\documentstyle[12pt]{article}
\begin{document}
\par\noindent
December 1995\\
LPTHE Orsay 95/83
\vspace{1cm}
\begin{center}
{\bf PERSPECTIVES IN LATTICE GRAVITY}\\
\vspace{0.8cm}
A. Krzywicki\footnote{Electronic mail address:
krz@qcd.th.u-psud.fr}\\
Laboratoire de Physique Th\'eorique et Hautes Energies, B\^{a}t. 211,
Universit\'e de Paris-Sud, 91405 Orsay, France\footnote{Laboratoire
associ\'e au C.N.R.S.}\\
\vspace{1cm}
{\bf Abstract}
\end{center}
\par\noindent
We briefly overview the development of Euclidean quantum gravity
in four dimensions regarded as a branch of statistical mechanics
of discretized random manifolds.
\vspace{1cm}
\par\noindent
{\em Contribution to a special issue of the Acta Physica Polonica
dedicated to W. Kr\'olikowski on the occasion of his 70th birthday}
\vspace{2cm}

\section{Introduction}
The absence of a fully consistent quantum theory of gravity is felt by
many theorists as a challenge. The corresponding research is not
motivated by present phenomenology. But it is likely that the difficulty
one encounters trying to merge together general relativity and quantum
mechanics reflects our misunderstanding of some basic issues and the
feeling that it might indeed be so is sufficient to trigger activity.
There exist several approaches to the problem. In this paper we shall
discuss only one of them.
\par
It is well known that the formulation of the classical theory
of gravity can start with the introduction of interacting tensor
fields living in a flat auxiliary space, provided one
imposes appropriate constraints, in the first place the gauge
invariance. The identification of the field with the metric of
the physical space-time is then done at the next stage.
It might appear that proceeding that way is the best strategy in
the quantization program: one has the correct classical theory at
the tree level, computing loops will give quantum corections.
Unfortunately one gets a theory plagued with infinities. Upgrading
the symmetry to super-symmetry does not resolve the problem. It
is by now commonly admitted that the basic objects should be
extended, which complicates the story considerably. In spite of
the appeal of super-string theories and of their potential
ability to unify interactions, it is fair to say that the
progress in this direction is slow if not uncertain.
\par
The approach to be discussed starts from an attempt to give a precise
meaning to Feynman's path integral over all metrics of a manifold. This
is, perhaps, closer to Einstein's geometrical intuition, since the
metric remains the central concept of the theory. The price to pay
is that one has to adopt the Euclidean version of the latter \cite{haw}.
Someone may object at this point that the equivalence of the
Euclidean and Minkowskian theories is questionable in the case
of gravity, which renders the whole approach suspect. Notice,
however, that Euclidean and Minkowskian gravities have in common
several of their salient features: gauge symmetry, perturbative
non-renormalizability, bottomless action. A successful
quantization of the former, which is doubtlessly more tractable,
is likely to be a prerequisite for the understanding of
the latter.
\par
Because of lack of space we shall leave aside most of
the developments concerning quantum gravity in less than
four dimensions ($4d$). We would like to stress, however,
that the study of $2d$ gravity (random surfaces) triggered
by Polyakov's famous paper \cite{pol} has to a large extent
inspired the research reviewed below. The following sections
express a personal view of the subject. Are left aside,
in particular, the papers where the dynamical triangulation
recipe (see sect. 2) has {\em not} been adopted
(see e.g. \cite{berg,ham2}).
We appologize to all those whose work has not been given due
attention.
\section{Discrete theory}
The pure gravity theory is defined formally by the path integral
\begin{equation}
Z = \int [{\cal D}g_{ab}] e^{-S}
\label{cz}
\end{equation}
\noindent
Here $S$ is the action, which is usually assumed to
have the Einstein-Hilbert form
\begin{equation}
S =  \int_{\cal M} d^dx \sqrt{g} \;
 (\Lambda - {1 \over  {16\pi G}} R \; )
\label{ca}
\end{equation}
\noindent
and ${\cal M}$ is a compact closed manifold. The integration involves,
in principle, a summation over all topologies and, for a given
topology, the integration over all metrics that can be obtained
one from another by a continuous deformation. Actually, for $4d$,
the summation over topologies is ill defined, and should
be restricted, say, to smooth manifolds. In $2d$,
where the classification of topologies is simple, the
number of manifolds grows like a factorial of the genus \cite{wein2},
so that the sum is not Borel summable. The situation in $4d$
is certainly not simpler. In practice,
most studies at $d > 2$ use a discrete formulation of the theory
and assume the topology is fixed.
\par
The discretization of a theory invariant under general
coordinate transformations is not a trivial matter. The
first basic idea is due to Regge \cite{reg}, who suggested
to replace the continuous manifold ${\cal M}$ by a collection of
flat $d$-simplexes forming a simplicial complex. The curvature
then resides on ($d-2$)-dimensional hinges. The second
important idea is that of {\em dynamical triangulations}
\cite{weing,dav,amb,kaz}: the simplexes are assumed to
be equilateral and the sum over metrics is replaced by
the sum over all possible manners to glue them together.
For an ensemble of exactly solvable models in $2d$ one can show
that the continuum and the discrete version belong to the
same universality class, when the dynamical triangulation
recipe is adopted.
\par
Let us denote by $N_k$ the number of k-simplexes in the
complex. For $d=3$ and 4 only two of these quantities are
independent. When
the simplexes are equilateral, the RHS of
(\ref{ca}) discretized \`a la
Regge \cite{ham} becomes a linear
combination of these two independent
numbers, which is remarkably simple. For $4d$ one can write
\begin{equation}
S = \kappa_4 N_4 -\kappa_2 N_2 ,
\label{dac}
\end{equation}
\noindent
where $\kappa_2 \sim a^2/G$ and $a$ denotes the lattice unit.
The partition function (\ref{cz}) takes the form
\begin{equation}
Z(\kappa_2, \kappa_4) = \sum_{N_2, N_4} Z_{N_2, N_4} e^{-S}
\label{dz}
\end{equation}
\noindent
where
\begin{equation}
Z_{N_2, N_4} = \sum_{T(N_2, N_4)} W(T)
\label{micro}
\end{equation}
\noindent
The sum is over all $4d$ closed manifolds
$T(N_2,N_4)$ of, say, spherical
topology, with fixed $N_2$ and $N_4$. The symmetry factor
W(T) equals the number of distinct labelings of the vertices
of $T$ divided by $N_0$!. The model is now defined precisely
enough to be converted into a computer code.
\par
The lattice can be further decorated with matter fields,
for example with Ising spins. This does not present any
conceptual difficulty. We limit ourselves here to pure
gravity, since for $d > 2$ the study of models involving
matter fields has not been pushed far enough to warrant
discussion in a short review.
\section{Numerical algorithms}
Any two combinatorially equivalent simplicial
complexes\footnote{For $d < 7$ triangulated smooth
manifolds of the same topology are combinatorially equivalent.}
can be
connected by a series of moves introduced long ago by
Alexander \cite{alex}. A smaller set of local moves
has been proposed in the context of lattice
gravity \cite{gg,am}. All these moves can be introduced as
follows: let a collection of $d$-simplexes
in a $d$-dimensional manifold be a part of the boundary
of a $(d+1)$-simplex. A move consists in replacing
the collection in question by the rest of the boundary.
Since $(d+1)$-simplex has $d+2$ $d$-dimensional faces,
there are $d+1$ possible moves of this type. Each move has
a reciprocal and, when $d$ is even, there is
one self-reciprocal move.
It has been pointed out in \cite{bk} that for $d=3$ all
Alexander moves can be constructed using these simple ones.
The formal proof valid for all $d < 5$ can be found in ref.
\cite{gv}. The ergodicity of the simple moves follows from
the fact that all the Alexander moves are reducible to
them\footnote{The moves used earlier
in the context of $2d$ gravity are identical or reducible to
these ones. However, the situation in $2d$ is particularly
simple and the ergodicity is proved by
rather elemenary methods.}.
\par
A word of caution is in order at this point: although
any two simplicial complexes can be deformed
one into another by a finite
number of simple local moves, the number of steps needed to
connect any two lattice configurations might grow so fast with the
volume that the ergodicity would not be insured
in practice. The possibility of this unpleasant scenario has
to be kept in mind.
\par
The further implementation of these ideas in computer software
has been greatly facilitated by the experience gained in
developing algorithms appropriate for random
surfaces \cite{kaz,adfo,jkp}. In this respect the techniques
worked out to simulate the so-called grand-canonical
ensemble\footnote{That is the ensemble where the
number of nodes is fluctuating.} of
random surfaces \cite{adfo,jkp}  are particularly instructive.
Anyone wishing to participate in the numerical studies of
quantum gravity is advised to start by getting conversant
with them.
\par
The efficiency of the algorithms is considerably improved
when the ergodic local moves are supplemented by the global
ones, where entire baby-universes are cut out at one place and
glued elsewhere \cite{abbjp,aj1} \footnote{These global moves are
not ergodic by themselves.}. We shall come to baby-universes
later on.
\section{Entropy of random manifolds}
In order for the theory to make sense the entropy of manifolds
should be an extensive quantity. In other words,
the number $Z(N_d)$ of distinct $d$-dimensional
simplicial complexes, made up of $N_d$ $d$-simplexes
and with fixed topology, should be bounded by $\exp{(c N_d)}$,
with $c$ being some finite positive number. It has been a surprise for
the physicists who got interested in the problem to learn that their
colleagues from the maths department have no idea how $Z(N_d)$ does
behave when $d > 2$ and $N_d \to \infty$.
\par
In $2d$ the exponential bound has been
proved analytically \cite{tut}. A numerical
evidence for such a bound in $3d$ has been given first in
ref. \cite{av} for spherical topology, and confirmed by
later studies. A similar result has been obtained for $d=4$ in
\cite{aj2}. There has been a controversy concerning the validity
of this result, but the present consensus is that it is correct (see
the review in \cite{cat}). Of course, numerical
evidence is not a proof. Hence,
several people presented analytic arguments to the effect that the
exponential bound does hold. It seems that these claims
rest on too restrictive assumptions, but we do not feel
expert enough in topology to develop this point.
\section{Phase diagram}
Let us take the existence of the exponential bound
discussed in the preceding section for granted :
\begin{equation}
\log Z(\kappa_2, N_4) \sim \kappa_{4crit} (\kappa_2) N_4 + ...
\label{zcan}
\end{equation}
\noindent
where
\begin{equation}
 Z(\kappa_2, N_4) = \sum_{N_2} Z_{N_2,N_4} e^{\kappa_2 N_2}
\label{zcan2}
\end{equation}
\noindent
and the subleading terms have not been written explicitely for
simplicity. It is clear from (\ref{dz}) that the theory does not
exist for $\kappa_4 < \kappa_{4crit}(\kappa_2)$.
As $\kappa_4$ approaches the critical
line $\kappa_4 = \kappa_{4crit}(\kappa_2)$ from above
the partition function $Z(\kappa_2,\kappa_4)$ develops a
singularity.
\par
Notice, that in pure quantum gravity it
does not make much sense to attach
physical significance to $\Lambda$ and $G$
separately. The content of
the theory, as defined  by (\ref{cz}) remains unchanged under the
rescaling of the metric $g_{ab} \to s g_{ab}$, which corresponds to
$\Lambda \to s^{d/2} \Lambda$ and $G \to
s^{-d/2+1} G$. Only the invariant
combination $\Lambda G^{d/(d-2)}$ is
relevant (see the discussion in \cite{kn}). In other
words one has to tune both $\kappa_ 4$
and $\kappa_ 2$ in order to
define the continuum theory. One needs for that a critical point
on the line $\kappa_4 = \kappa_{4crit}(\kappa_2)$.
\par
Such a point has first been discovered in ref. \cite{bk}, in
the context of $3d$ gravity (which in this respect resembles the $4d$
one, except for the order of the transition, see later). It has been
found that below that point the system is in a crumpled phase, where
the average number of nodes per simplex tends to zero when the number
of simplexed is sent to infinity. Above the critical point this ratio
tends to a finite limit, so that at least a sensible thermodynamical
limit can be defined. An analogous critical point has subsequently
been found for $d=4$ \cite{aj3,am2}. Contrary to $3d$, where it is
of first order \cite{abkv}, the transition in $4d$ appears to be
continuous \cite{aj3,ckr,aj1}.
This is what one might hope, since
in $4d$ there should be a place for the graviton, absent in $3d$.
It is now customary to refer to the crumpled phase as to the {\em
hot} one. The phase above the critical point is called {\em cold}.
A careful analysis
\cite{aj1} of the phase structure in $4d$ further demonstrates
that the {\em internal} fractal dimension $d_H$ of the manifolds
is close to 2 in the
cold phase and presumably infinite in the hot one.
\par
It is worth mentioning at this place that the latice theory always
has a well defined most probable configuration (vacuum state). It
appears that this vacuum is not just the state with largest
curvature, which on a dynamically triangulated manifold is necessarily
finite. The vacuum seems to be nontrivial and is stabilized by the
entropy od manifolds, which in this formulation
of the discrete theory is
defined unambiguously \cite{bk,abkv}.
\section{Baby universes}
Baby universes (BU) are sub-universes connected to the rest of the
universe by a narrow neck. They are Euclidean analogs of black
holes and early speculations concerning Euclidean quantum gravity
have already introduced this concept \cite{haw}. The numerical
simulations of random manifolds have revealed that the emergence of
BU is an extremely common phenomenon, in all
dimensions that have been considered. It is very unlikely that a
random manifold remains more or less smooth
(in the intuitive sense of the word). If one starts a simulation
with a smooth manifold, soon there are BU growing out
of it. Further, there are BU growing on BU
and so on. The final structure is tree-like. It
can be demonstrated analytically in $2d$ \cite{jm}
that this tree is a fractal. In $4d$ the tree-like structure is
especially manifest near and above the critical point. Actually, in
the cold phase, the tree resembles a branched polymer \cite{aj1}.
The tree has the topology of a sphere because the algorithm keeps
the topology fixed by construction. It is very likely that
the typical configuration would remain a collection of sub-universes
connected by wormholes if one succeeded to upgrade the algorithm so as
to allow the topology to change. The possible relevance of such
a geometry for the cosmological constant problem has been pointed out
long ago by Hawking, Coleman and others \cite{haw}.
\par
The discovery of the tree-like geometry of typical random manifolds
with fixed topology is a very important and a very intriguing finding.
A quantized manifold does not resemble at all the familiar systems
making small quantum fluctuations around a smooth clasical
configuration. It is perhaps not surprising that the construction
of the quantum space-time starting with interacting elementary
entities (see the Introduction) is not a simple matter.
\par
The average number $n(N_B, N_4)$ of
BU with a given volume $N_B$ can be found using
a combinatorial argument
\cite{jm}
\footnote{The neck of a BU can be
regarded as a puncture on each of the two
parts of the manifold it connects.}. The result
is particularly simple when one assumes that
\begin{equation}
Z(\kappa_2, N_4) \sim N_4^{\gamma-3} e^{\kappa_{4crit} N_4} ,
\label{zcan3}
\end{equation}
\noindent
which is true in $2d$ and is likely to
hold in $4d$ in the
vicinity of the critical point:
\begin{equation}
n(N_B, N_4) \sim N_4 [(1 - {{N_B} \over {N_4}})
 N_B]^{\gamma - 2} \; , \; N_B < N_4
\label{buvol}
\end{equation}
\noindent
There exist general arguments \cite{dfj}\footnote{Strictly
speaking, this paper deals with random surfaces. However
the geometrical arguments
employed are certainly of more general validity.}
to the effect that generically $\gamma \leq 0$ or
else the manifolds degenerate into branched polymers with
$\gamma = {1 \over 2}$.
Thus the educated guess is that in the sensible
sector of the theory the
number of BU carrying a {\em
finite} fraction of the total volume is $\sim N_4^\gamma$
and tends to a constant or vanishes
when $N_4 \to \infty$, i.e. in the continuum limit.

\section{Scaling and renormalization group}
Recently, much activity has concentrated on
the behavior of the discrete theory in
the neighbourhood of the critical point. We have no
place here to give justice to all this effort and, in
particular, to all the facets of the particularly
thorough work by Ambj\o rn and Jurkiewicz \cite{aj1}
(we have already refered to it on several occasions).
\par
The geometry of the ensemble of manifolds can be
characterized by invariant correlations between
local operators $O(x)$. The simplest correlator is the
two-point function with $O(x) = 1$. On a lattice
it takes the form
\begin{equation}
G(r, N_4, \kappa_2) =  N_4^{-2} \langle
\sum_{A,B} \delta (r- \mid
 x_A-x_B \mid) \rangle_{N_4}
\label{pcf}
\end{equation}
\noindent
where $\mid x_A-x_B \mid$ is the geodesic distance
between simplexes $A$ and $B$. The large distance
behavior at fixed $\kappa_2$ is \cite{aw,aj1}
\begin{equation}
G(r, N_4, \kappa_2) \sim e^{-c(r/N_4^{1/d_H})^{{d_H}
 \over {d_H-1}}}
\label{pcf2}
\end{equation}
\noindent
where $c$ is some constant and $d_H$ is the internal
Hausdorff dimension. Both can depend on $\kappa_2$. For
large enough $N_4$
\begin{equation}
\langle r \rangle_{N_4} \sim N_4^{1/d_H}
\label{avr}
\end{equation}
\noindent
The reciprocal relation
\begin{equation}
\langle N_4 \rangle_r \sim r^{d_H}
\label{avN}
\end{equation}
\noindent
also holds \footnote{The LHS is the average number of simplexes in
manifolds with two boundaries separated by invariant geodesic
distance $r$.}. The finiteness of $d_H$ has been assumed. The
scaling manifest in (\ref{pcf2}) has been earlier observed
empirically \cite{dbs} in the full range of $r$:
$G(r, N_4)$ is mostly a function of $r/\langle r \rangle$. It
has been further claimed in \cite{dbs} that this function has
an approximately constant shape along trajectories in
$(N_4, \kappa_2)$ plane.
\par
It is tempting to attack the problem of scaling
using the techniques of the real space renormalization group.
The very definition of a blocking procedure is non-trivial
in this context: ideally, the blocking should be a self-similarity
transformation, a constraint difficult to satisfy when one deals
with a random lattice. It has been
proposed in \cite{jkk} to define
the renormalization group (RG) transformation
as the process of cutting the
last generation of baby universes, that is those
BU which have no further BU growing on them \footnote{In
practice, one cuts only the minimum-neck BU (minBU), which are
easy to identify.}. Under this operation the tree
gets smaller, in lattice units, and less branched, which is
interpreted as reflecting the loss of the resolving power.
\par
Let us keep fixed the {\em physical} volume $V =N_4 a^4$ of
the manifold. Consider the moments $\langle r^k \rangle$ of
the correlator (\ref{pcf}). They transform under RG:
$\langle r \rangle \to \langle r \rangle - \delta r$, etc.
Assuming that $\kappa_2$ is the only coupling relevant for
the in-large geometry one has along the RG flow
\begin{equation}
\delta r = r_N \delta \ln{N_4} + r_\kappa \delta \kappa_2
\label{step}
\end{equation}
\noindent
where $r_N$ and $r_\kappa$ are the partial derivatives of
$\langle r \rangle$ with respect to $\ln{N_4}$ and $\kappa_2$,
respectively. Furthermore,
\begin{equation}
\delta \ln{1 \over a} = {1 \over 4} \delta \ln{N_4}
\label{step2}
\end{equation}
\noindent
{}From (\ref{step}) and (\ref{step2}) and using computer data
one can calculate the $\beta$-function \cite{bkk,bbkp}
\begin{equation}
\beta(\kappa_2) = {{d \kappa_2} \over {d \ln{{1 \over a}}}}
\label{bet}
\end{equation}
\noindent
It is found that the theory posseses an ultra-violet stable
fixed point $\kappa_2 = \kappa_{2crit}$. The value of
$\kappa_{2crit}$ obtained from RG is close to
that found by other methods.
Thus, in the neighborhood of the critical point
\begin{equation}
\beta(\kappa_2) = \beta_0 (\kappa_{2crit} -  \kappa_2)
, \; \; \beta_0 > 0
\label{beta}
\end{equation}
\noindent
Integrating (\ref{bet}) one gets
\begin{equation}
a = a_0 \mid \kappa_{2crit} -  \kappa_2 \mid^{1 \over {\beta_0}}
\label{latstep}
\end{equation}
\noindent
where $a_0$ is an integration constant, which should be given a
value, in physical units, in order to define the theory.
The RG flow lines are
\begin{equation}
N_4^{\beta_0/4} \mid \kappa_{2crit} -
  \kappa_2\mid \; \equiv t = V/a_0^4
\label{rgflow}
\end{equation}
\noindent
The continuum limit is
\begin{eqnarray}
N_4 \to \infty \nonumber \\
\kappa_2 \to \kappa_{2crit} \nonumber \\
t = {\rm const}
\label{cont}
\end{eqnarray}
These results are closely analogous to
those obtained in $2+\epsilon$
dimensions in the continuum framework (see \cite{kn} and
references therein). It follows from the above discussion that
one should be careful in interpreting results obtained at
fixed $\kappa_2$: the line $\kappa_2 =$ const intersects an
infinity of RG trajectories, each
representing a distinct version
of the theory.
\par
Other interesting correlators are those obtained setting
$O(x) = R(x)$, where $R(x)$ is the scalar curvature. Their
integrals are unambiguously defined:
\par
\begin{equation}
m_k(N_4) = {{\partial^k \ln{Z(\kappa_2, N_4)}}
\over {\partial \kappa_2^k}}
\label{rrcor}
\end{equation}
\noindent
and are the cumulants of the node distribution\footnote{One
has $N_2=2(N_4+N_0-2)$.}. Computer
data \cite{aj3,ckr,var,
bbkp} are compatible with simple
finite-size scaling
\begin{eqnarray}
m_2(N_4) \sim N_4^b f[(\kappa_2 - \kappa_{2crit}) N_4^c] \\
m_3(N_4) \sim N_4^{b+c} f'[(\kappa_2 - \kappa_{2crit}) N_4^c]
\label{fss}
\end{eqnarray}
\noindent
This suggests the existence of a finite mass gap
scaling to zero when
$ \kappa_2 \to  \kappa_{2crit}$.
As long as one is on the lattice the mass
gap {\em is expected} to be finite,
since the continuum gauge symmetry responsible for the
existence of the graviton is absent. According
to the preliminary data \cite{bbkp}
$\beta_0/4 > c$. This seems to indicate that
the mass gap vanishes in the
continuum limit (\ref{cont}) as it should. Much
more work will be necessary to check the spin of the corresponding
particle.

\section{Conclusion}
  Let us conclude with a few words about the open
questions. There are, of course, the fundamental questions
relative to the summation over topologies, the continuation to
real time etc, which require very brigth new ideas. There are
also more accessible problems, within the lattice formalism.
The central one is the nature of the continuum
limit and, in particular, the search of an
evidence for the graviton. The next
one is a careful study of the interaction of matter fields
with geometry. Is all this sensible
and does it for sure correspond to a genuine
gravity theory, let it be Euclidean ? What is certain is that
the progress in this field is rapid and that people involved
 have a lot of fun!

\end{document}